\documentclass[a4paper]{article}

\usepackage[english]{babel}
\usepackage[utf8x]{inputenc}
\usepackage[T1]{fontenc}

\usepackage[a4paper,top=3cm,bottom=2cm,left=3cm,right=3cm,marginparwidth=1.75cm]{geometry}

\usepackage{array,multirow}
\usepackage{xcolor,colortbl}
\usepackage{amsmath}
\usepackage{graphicx}
\usepackage[colorinlistoftodos]{todonotes}
\usepackage[colorlinks=true, allcolors=blue]{hyperref}
\usepackage{rotating}
\usepackage{footnote}
\definecolor{red}{rgb}{0.898,0.4509,0.4509}
\definecolor{green}{rgb}{0.6823, 0.835, 0.505}
\makesavenoteenv{tabular}
\makesavenoteenv{table}
\makeatletter
\newcommand\footnoteref[1]{\protected@xdef\@thefnmark{\ref{#1}}\@footnotemark}
\makeatother

\title{To Agile, or not to Agile:\\ A Comparison of Software Development Methodologies}
\author{Ruslan Shaydulin, Justin Sybrandt}

\begin{document}
\maketitle

\begin{abstract}
   Since the Agile Manifesto, many organizations have explored agile development methods to replace traditional waterfall development. Interestingly, waterfall remains the most widely used practice, suggesting that there is something missing from the many ``flavors'' of agile methodologies. We explore seven of the most common practices to explore this, and evaluate each against a series of criteria centered around product quality and adherence to agile practices. We find that no methodology entirely replaces waterfall and summarize the strengths and weaknesses of each. From this, we conclude that agile methods are, as a whole, unable to cope with the realities of technical debt and large scale systems. Ultimately, no one methodology fits all projects.
\end{abstract}

\section{Introduction}

Agile software development, a term introduced in 2001 in famous Agile Manifesto \cite{agilemanifesto}, took the world by a storm and quickly became, if not the most popular then certainly the most fashionable, development style. However, there is a lot of misunderstandings surrounding the agile development style. As teams adopt and change this methodology they raise questions about the limitations \cite{turk2014limitations} and applicability \cite{qumer2008evaluation} of this approach.

In this paper we explore the state-of-the-art methodologies which are widely used by software companies today with a focus on how agile methods have been altered to facilitate the need for planning. We compare different software development practices, both agile and traditional, in an attempt to find an answer to the questions like "Why does the waterfall persist?" and "What makes popular agile methods popular?"

This report is structured in the following way: 
Section 2 consists of the historical idealogical foundation of agile development, describing of the most famous agile method, extreme programming, and a quick discussion of architectural agility.
Section 3 describes the evaluation methodology that we used to compare various methods and the rationale behind our evaluation framework.
Section 4 contains descriptions of each studied methodology as well as short discussions on how they function in real-world circumstances. This section includes the evaluation results of seven of the most popular software development methods.
Finally, ins section 5 we draw conclusions and outline possible future work.

\section{Agile}

In this section we will describe the original agile approach including its origins and its current state. This will act as a baseline on which we will compare modern reformulations of these core principles. Furthering this baseline we discuss Extreme Programming and Architectural Agility. The former is the principle agile methodology, the latter is a more modern approach with attempts to incorporate architectural design into the process. Although neither method is extraordinarily popular today, these serve to bookend the extremes of the agile development described in latter sections.

\subsection{Agile: Basics}

Martin, one of the authors of the original Agile Manifesto, begins his book on Agile Software development by defining agile as ``the ability to develop software quickly, in the face of rapidly changing requirements'' \cite{martin2003agile}. This approach is motivated by what he calls the ``runaway inflation'' of the software development processes. This inflation begins when a team fails to deliver a product on time, and begin over-engineering a methodology to attempt to improve their performance. If the new methodology also leads to a failed project, an even more complicated replacement may be put in place. From there the cycle continues. As a result of this inflation, many companies were stuck in a quagmire of increasingly complex processes. A group of industry experts met and decided to create a set of values and principles that would help companies overcome these problems calling themselves the ``Agile Alliance'' \cite{alliance2001agile}.

There are many methods developed around the principles described in the agile manifesto, and it is important to separate the principles of agile from the methods and practices of agile software development (e.g., SCRUM). Failing to make this separation often leads to blindly following the word of practices of a certain method and ignoring the spirit of them. This can result in dangerous and inefficient cargo-cult mentality and in failure to deliver a good product. As a baseline methodology, we begin by exploring extreme programming (XP) before moving onto the seven most popular modern design methods.

\subsection{Extreme Programming}

Extreme programming is a method that is built around customers interacting closely with the development team throughout the course of the project \cite{beck2000extreme, martin2003agile}. Customers see the system come together. As they learn more about what they need, customers and developers work together to adjust system requirements. To incorporate this flexibility, instead of producing detailed requirements, customers create user stories. These are simply a few words on an index card that summarize the rough idea of a requirement. As the system comes together, the customer sees the progress of the project and adjusts his or her expectations accordingly. 

Development in XP moves in short cycles, and at the end of each iteration a system is shown to the customers in order to get their feedback. The average cycle lasts only two weeks long and consists of writing just enough code, typically in pairs, until a set of tests pass \cite{beck1999embracing}. Speed is paramount, and code is integrated several times a day. However, the project is treated as a marathon and not a sprint, and therefore developers are not allowed to work overtime \cite{beck2000extreme}.

XP purposely ignored the practice of architectural design, and instead insists upon a minimally viable architecture. It is considered essential to avoid unnecessarily complex designs and resist adding infrastructure before it is strictly needed \cite{paulk2001extreme}.

Workload is managed in XP through a series of user stories and story points \cite{beck2001planning}. At the beginning of each iteration a set of stories is created. Then they are assigned points that measure the cost of their implementation. The stories are broken into tasks that should take 4-16 hours for a developer to implement, after which each developer gets to pick tasks according to his or her budget. These budgets are determined by how many story points each developer completed during the last iteration. In a similar fashion, the total cost of user stories is determined by the number of story-points completed on the previous iteration. In this manner, XP is able to iteratively refine the cost and time estimate for a complex software project.

\subsection{Architectural Agility}
Enabling Agility Through Architecture

Brown et al. describe a method to produce ``just-in-time'' architecture designs while following an agile development process \cite{brown2010enabling}.
In their new method, "Architectural Agility," they modify the typical agile iteration process two introduce some forward-thinking elements.
In an effort to improve the ``Enhancement Agility,'' the speed at which a software team can introduce new features, Brown introduces project introspection to the agile iteration feedback loop. In a sense, this expands upon the iterative budget refinement present in XP, but focuses on technical debt.
While practicing architectural agility, a team should conduct real options analysis and technical debt management at the start of each new iteration.
This process begins with a technical research phase wherein potential future features are proposed and explored.
This task is only intended to quickly identify routes the project might soon take, and produces a list of potential features and the architectural components they depend on.
Then, real options analysis evaluates both the cost of taking actions in the upcoming iteration to facilitate those new features, as well as the potential expense of putting off the decision.
Afterwards, technical debt management involves determining the cost of making modifications to the current codebase without major refactoring.

The end goal of architectural agility is an increased focus on the dependencies between user stories. This begins with an additional category added to the traditional agile release planning board, and ends with a dependency structure matrix.
By focusing on these dependencies, as well as quality attributes, a team should be able to quickly identify the software components which are most central to the proposed system.
This allows a team to understand which elements require fleshed-out architectures in the upcoming iteration.

Ultimately, it would seem that the project management structure proposed by Brown et. al. attempts to find a middle ground between the speed of agile, and the patience of waterfall.
By taking the time each iteration to look a couple steps ahead, this method should be able to improve agile teams with complex software architectures.

\section{Evaluation Methodology}
\label{sec:evalMethod}

In the following sections we detail the basis for our methodology evaluation and go on to explain our approach and the specifics of our evaluation criteria.

\subsection{Literature review}

In \cite{kitchenham1996evaluating} Kitchenham describes the DESMET evaluation methodology. This method focuses on software development from multiple angles, taking input from surveys, case studies, and actual experiments. Both quantitative and qualitative results can be extracted from these three data sources. For example, case studies can both show the exact process as well as the opinions of experienced developers. Unfortunately, the DESMET method only works when a department is able to control their development process to ensure valid results, and therefore, makes the methodology difficult to apply in many agile environments. Still, we can borrow a similar data gathering approach our own evaluation methodology.

In \cite{sorensen1995comparison}, Sorensen compares waterfall software development models with incremental ones such XP and other agile methods. Though he shows many similarities, such as the fact that both allow for work force specialization, Sorensen highlights the ability of incremental methodologies to start out without clear set of requirements, as opposed to waterfall which requires the full set of requirements at the outset. Additionally, he notes that incremental methods require clean interfaces between modules while the waterfall approach is capable of maintaining more complex module connections.

In \cite{thayer1981major} Thayer, Pyster, and Wood detail twenty problems that many software projects face during development. Although this work was published in 1981, it is clear that most issues found on this list still plague a plethora of modern projects. For instance, these questions focus on topics such as unclear project requirements, poor project quality, and poor project estimates. We also note that while the first ten questions focus on the software itself, it remaining ten focus on a variety of important tangential topics, including management and testing. This, to us, highlights the importance of addressing non-functional qualities of the software creation process.

In \cite{awad2005comparison}, Awad compares what he calls ``light'' agile methods with ``heavy'' waterfall-like methods. For example, he points out that heavy approaches favor a ``Command-Control'' culture, whereas agile approaches tilt toward a culture of ``Leadership-Collaboration''. The same question of requirements, discussed in \cite{sorensen1995comparison}, is described in terms of ``minimal upfront planning'' and ``comprehensive planning.'' In his work, Awad attempts to describe the balance between both approaches and shows that agile works well for projects that allow for inexpensive refactoring.

\subsection{Our approach}

Following approach introduced by Awad, we attempt to find a balance for the methodologies we encounter. In order to do so, we extend the set of project characteristic discriminators, introduced by \cite{awad2005comparison}. Additionally, we adapt and utilize the questions asked by Thayer et. al. in order to evaluate how different software methodologies and architectural frameworks address a common set of concerns \cite{thayer1981major}. We note that the exact wording of these questions has, in some instances, gone out of date. Others simply lie too far outside the scope of a software development methodology to be applicable. For these reasons, we restrict the set of initial questions to a subset of topics. These topics are explained in detail below. For each evaluated methodology, we describe if and how it addresses these nine topics. We hypothesize that many will not strictly bind a development team to a specific organizational structure or design selection, but instead will instead influence these topics in an indirect manner. For example, traditional agile with scrum \cite{schwaber2002agile} tends to produce very ``wide'' team hierarchies where all developers have very similar standings when making decisions. Additionally, we evaluate the agility of each methodology in an effort to identify if a strong correlation exists between software quality and agility. We do so in a similar manner, also described in the following section.

\subsection{Methodology Description}

Using sources and methods described above, we created the following evaluation methodology matrix. We define our own set of criteria which any good software development strategy should fulfill.

\subsubsection{Process Quality Evaluation Criteria}
\begin{enumerate}
\item \textbf{Requirements flexibility}: the method is able to respond to new and changing requirements.
\item \textbf{Requirements fulfillment guarantee}: method contains a mechanism to guarantee functional and non-functional requirements in order of their importance.
\item \textbf{Cost estimation}: the method has the ability to estimate the resources required to complete specific development tasks.
\item \textbf{Cost estimates refinement}: the method is able to update these cost estimates over time.
\item \textbf{Validation}: the method is capable of identifying when tasks are done properly vs when tasks are incomplete or contain errors.
\item \textbf{Quick validation}: the method is capable of quickly detecting when non-functional requirements are violated.
\item \textbf{Focus on customer}: the method focuses on client needs and ensures they are met continuously over the lifetime of the project.
\item \textbf{Understandability guarantee}: the method encourages the development of an understandable and maintainable code base.
\item \textbf{Technical debt control}: the method prevents runaway inflation of the cost of new features and bug fixes while controlling the amount of technical debt.
\end{enumerate}

In addition to our quality criteria, we also present a series of criteria for evaluating the agility of a given software method. We do so in order to detect if there is a measurable correlation between method quality and method agility. This series of questions is directly inspired from the set of project characteristic discriminators defined in  \cite{awad2005comparison}. 

\subsubsection{Process Agility Evaluation Criteria}
\begin{enumerate}
\item \textbf{Prioritizes added value} The method prioritizes added value over nonfunctional benefits, refactoring, or documentation.
\item \textbf{Allows partial requirements} The method does not require stable or fully defined requirements.
\item \textbf{Focuses of small teams} The method emphasizes small ``wide'' teams over large hierarchical ones.
\item \textbf{Develops minimal viable architecture} The method does not develop software architecture beyond what is needed for the current working set of software requirements.
\item \textbf{Produces minimal documentation} The method does not produce much documentation, and the documentation which is produced is not intended to last the lifespan of the project.
\item \textbf{Relies heavily on customer feedback} The method includes customers at multiple points in the design process, as opposed to a single requirement gathering phase.
\item \textbf{Susceptible to unforeseen risks} The method is weak to unforeseen major risks or complications which were not foreseen. Waterfall, for example, spends substantial time documenting these risks and is therefore less susceptible.
\end{enumerate}
\section{Evaluated Methods}
\label{sec:methods}

Vijayasarathy et. al. in \cite{vijayasarathy2016choice} conducted a survey in which they asked 153 developers to describe their software development process. One of the most important results of this study is a listing of the most commonly used software development practices as of 2016. We use these results to inform our methodology reviews, allowing us to review the most widely used methodologies. The list of software methodologies which address in later sections follows the same order of popularity.

\begin{enumerate}
\item Waterfall
\item Agile Unified Process
\item Scrum
\item Test-Driven Development
\item Rapid Application Development
\item Joint Application Development
\item Feature-Driven Development
\end{enumerate}

\subsection{Waterfall}
In \cite{royce1970managing}, Royce describes the process which would eventually become known as waterfall development. This method includes seven steps which are completed in order and should result in stable software as a result. These steps include, system requirements, software requirements, analysis, program design, coding, testing, and operations.  Royce notes that these steps do not typically form a linear flow from one to the next, instead many backward pointing edges between these states may exist. For example, as testing reveals bugs, a project might need to return to the program design phase. Additionally, these main phases represent sub-trees within them selves. For example, Royce points out that whole departments might be assigned to just the testing phase.

Waterfall typically requires a long time of requirements gathering and project planning before any code is written. This process attempts to identify the full set of project requirements. Other methods, such as \cite{benington1983production} by Benington  and \cite{ramamoorthy1985metrics} by Ramamoorthy also incur a large up front requirement gathering cost and were published around time same time. Since then, numerous other flavors of waterfall development have been introduced such as the iterative model described in \cite{munassar2010comparison}, which closely resembles the ``iterative waterfall method.''

It is unlikely that those survey respondents who reported using waterfall in 2016 followed the original specification without modification  \cite{vijayasarathy2016choice}. Waterfall has changed a lot over the years as developers have began experimenting with a more iterative approach. For the purposes of our analysis, we consider all of the above waterfall variants together, including iterative agile. As a whole, these methods are characterized by large upfront costs in requirement gathering, inexpensive bug fixing, accurate  time estimates, and inflexibility to changing or uncertain requirements. It seems clear that there are still many projects which are sufficiently well-defined to facilitate such a starched methodology.

\subsection{Agile Unified Process}

Agile Unified Process was developed by Scott Ambler \cite{ambler2002agile} as a simplified version of the Rational Unified Process. AUP development was stopped in 2006. In 2009 Disciplined Agile Delivery was introduced to supersede it, yet the survey results in \cite{vijayasarathy2016choice} make it clear that AUP remains very popular.

AUP applies agile techniques, such as Test Driven Development (TDD) and Agile Model Driven Development (AMDD) to improve the productivity of a team. It acts as a unifying framework for multiple agile methodologies and processes.

The AUD approach is ``serial in the large'' and ``iterative in the small''. It consists of four subprocesses or work flows: Modeling, Implementation, Testing and Deployment, each of which goes through four phases: Inception, Elaboration, Construction and Transition.

The principles of AUD emphasize simplicity. The simplest possible tools are preferred to complex products, minimal viable documentation is preferred to detailed documentation. It's aiming to prioritize high-value activities over trying to define everything that can possibly happen in a lifespan of a project.

Success of Agile Unified Process over other agile methodologies can be attributed to its flexibility, as well as to its simplicity.

\subsection{Scrum}

Scrum \cite{schwaber2002agile} is one of the most popular frameworks for implementing agile. Its defining characteristic is commitment to short iterations of work.

In scrum, a product is developed through a series of fixed-length iterations, called sprints, that allow for software updates at a regular cadence. The part of scrum that is the most attractive to a team is the idea that some call ``continuous inspiration'' \cite{schwaber2002agile}. Team members are motivated by tangible, visible progress at the end of each sprint, as well as by the ability to "show off" during the sprint demo.

Sprint consists of four ``ceremonies'' \cite{alliance2016learn}:
\begin{enumerate}
\item Sprint planning -- team meeting where the next sprint is outlined.
\item Daily stand-up -- daily 15 minute meeting for the team to sync up.
\item Sprint demo -- weekly meeting where teammates showcase what they will ship during the week.
\item Sprint retrospective -- weekly analysis of what went wrong and what went right during the previous week.
\end{enumerate}

Scrum has three specific roles: product owner, scrum master, and the development team. The product owner works with the business requirements and gives requirements to the team. The scrum master coaches the team and make sure the team observes scrum practices. The development team works closely together in an open and collaborative manner, talking regularly at scrum meetings. A potential point of failure, of course, comes from the quality of the product manager who is the sole point of contact between the development team and the customers.

\subsection{Test Driven Development}

Test Driven Development (TDD) is a design process which focuses on developing tests for ``units' before the implementation of the functionality in those units \cite{janzen2005test}. This practice creates a sort of programmers metacognition, where a significant portion of the development process centers on analyzing code deliverables. In relation to other methodologies discussed in this section, TDD relies the most heavily on specific testing technologies. For example, TDD would not be where it is today if it were not for automatic testing frameworks such as JUnit and ANT. Additionally, TDD often plays a secondary role in other agile methodologies such as Extreme Programming \cite{beck2000extreme}. This is primarily because TDD alone fails to cover the entire software design process, and instead focuses almost entirely on the technical aspects of writing code. After-all, how would one write a test to guarantee a quality attribute such as ``discoverability'' or ``flexibility.'' Nevertheless, TDD has recently grown into its own software design method \cite{damm2005introducing}. 

In \cite{maximilien2003assessing}, IBM employees address the efficacy of TDD in a large scale software system, JavaPOS. The IBM group noticed a higher defect rate than desired, so switched to a TDD-centric approach in an effort to improve their product. They did some initial design work with UML diagrams and a prototype, which they were able to test against the existing JavaPOS standard. As a result of this experiment, IBM noticed a 50\% reduction in error rate, a slight decline in productivity, and a more flexible design process.

A more recent study \cite{bissi2016effects} looks to explore the effect TDD has in the modern software production process. By aggregating over a thousand papers evaluating TDD, Bissi et. al. note that most TDD studies report an increase in code quality, and a decrease in productivity. Interestingly the study also found that almost every TDD project was written in C++ or Java. This speaks to the limited scope of TDD, and how tied it is to specific testing frameworks.

\subsection{Rapid Application Development}

Rapid Application Development (RAD) emphasizes user involvement in every step of the design process and was initially proposed by Martin in \cite{martin1991rapid}. In this method, software design occurs in four distinct phases: the requirements planning phase, the user design phase, the construction phase, and the cutover phase. These phases often follow the typical systems development life cycle \cite{jacobson1999unified}. In the requirements fathering phase all stakeholders, especially key users, are brought together to determine system requirements, like many other methodologies discussed here. What RAD does differently, is that user interaction continues into the design phase. As opposed to traditional agile methods, projects following the RAD method will prioritize prototyping over partially-functional deliverables. This means that prototypes can emerge as early as architecture planning time. By doing so, RAD ensures that user feedback can be incorporated as the system early and continuously. Additional benefits such as risk control, and budget sensitivity have also been noted as side-effects to the heightened sense of user involvement \cite{beck2000extreme}. Although it has also been noted that the rapid pace of prototyping can result in poor design as developers seek short term functionality while ignoring technical debt \cite{gerber2007practical, beynon1999rapid}.

A notable and interesting result of this methodology is the quantification of the user. By giving users so much power in the software design process, it can become ambiguous how to proceed when users inevitably disagree. Those practicing this methodology typically separates users into various roles, one of the most common being that of \textit{visionary} (one who instigates the project), \textit{ambassador} (one who represents the user community at large), and \textit{advisor} (one who is asked to influence design decisions) \cite{mackay2000reconfiguring}. Although such an approach can clarify many typical crossing points, arguably this highlights another weakness in the RAD development process. RAD results seem to be limited by the quality and commitment of the users a software team has access to.

\subsection{Joint Application Design}

Joint Application Design (JAD) is a methodology for operationalizing user involvement and user participation\cite{carmel1993pd}. Is assumes that the final quality of the software is proportional to the degree of final user involvement in the software's development\cite{hirschheim1991symbolism}. While some argue that this notion is unsupported and lacks any evidence \cite{ives1984user}, there is a clear intuition that the best way to understand the user's needs is to involve him or her in the development process and adjust the requirements as user's objectives change. JAD takes this approach to the extreme.

JAD is centered around a structured meeting, called the ``session,'' which the entire software development process resolves around. During the session, users and software developers discuss a clearly formulated agenda in an organized fashion. In some approaches the meetings are less structured and prefer the ``free discussion'' approach, however they have proved to suffer from a number of problems\cite{davidson1993exploratory}. This effort maintains contact between the development team and software end users throughout the lifespan of the software process.

JAD meetings have proved to be superior to interviewing techniques employed by other methods \cite{becker1993integrating} and is widely used in industry and even attracted the interest of the scientific community.

It is important to point out that JAD is nothing more than a technique for getting the requirements from users. It can, for example, be integrated into the Agile Unified Process. It, by no means, is a full substitute for a method like waterfall; it has to be complemented with a way to formalize the process of the actual application development.

\subsection{Featured Driven Development}

Feature Driven Development (FDD) follows along with many other agile methods in that it focuses on short value-adding sprints and attempts to introduce new working functionality in each iteration \cite{hunt2006feature}. What sets feature driven development apart is the speed and granularity of each sprint \cite{benoit1999feature}. Each feature is described as a short sentence containing a clear action, result, and object. For example a feature may be as short as ``display the shopping cart to the customer.'' These features are grouped into sets, and each set describes a two week iteration  \cite{palmer2001practical}. FDD promotes very modular architectures, as each feature should, in a perfect world, be concurrently developed.

Typically, FDD start with UML modeling. As features are planned, the necessary software components and their connections should grow organically. Arguably, this is the opposite of typical architectural design processes where a large system is split into encapsulated submodules. Additionally, FDD defines a hierarchy of developers, appointing some to be ``chief programmers'' responsible for small teams and planning sprints \cite{awad2005comparison}. These leaders act as a way to fluidly create and dismantle teams around features, as well as to regulate the quality of software produced in an iteration. Interestingly though, as opposed to many other agile methods, FDD actively discourages refactoring. Instead the development cycle focuses on added value to an extreme. After-all once a feature is present in the software, the customer is satisfied \cite{khramtchenko2004comparing}.

It would seem that FDD is a very natural, albeit naive, approach to software design. If a team is good and spends adequate time each iteration on the closing code review, it is possible that code quality may not deteriorate too quickly. It does appear that the method requires substantial upfront cost in the FDD version of requirements gathering in order to put together an initial plan that will not change too drastically during the lifetime of the project, otherwise an FDD project would require deep refactoring. Lastly, FDD fails to account for many non-functional qualities which typically depend on interactions between features. For example, FDD could easily produce a system had to both produce a shopping cart as well as track purchasing data, but would be ill equipped to ensure a latency requirement between user actions and the tracking system.

\bigskip
\begin{table}[h!]
\begin{center}
\begin{tabular}{ c  r | c | c | c | c | c | c | c |}
& & \multicolumn{7}{c|}{Methodology} \\\cline{3-9}
& & Waterfall & AUP & Scrum & TDD & RAD & JAD & FDD\\\hline 
\multirow{9}{*}{\rotatebox{90}{Quality Criteria}}
& Requirements flexibility 
	& \cellcolor{red}No
    & \cellcolor{green}Yes 
    & \cellcolor{green}Yes
    & \cellcolor{green}Yes
    & \cellcolor{green}Yes
    & \cellcolor{green}Yes
    & \cellcolor{red}No\\ 
& Requirements fulfillment guarantee 
	& \cellcolor{green}Yes 
    & \cellcolor{green}Yes 
    & \cellcolor{green}Yes 
    & \cellcolor{red}No \footnote{Only Functional}
    & \cellcolor{green}Yes
    & \cellcolor{red} No
    & \cellcolor{green}Yes\\
& Cost estimation 
	& \cellcolor{green}Yes 
    & \cellcolor{green}Yes 
    & \cellcolor{green}Yes 
    & \cellcolor{red}No
    & \cellcolor{green}Yes
    & \cellcolor{red}No
    & \cellcolor{green}Yes\\
& Cost estimates refinement 
	& \cellcolor{red} No 
    & \cellcolor{green}Yes 
    & \cellcolor{green}Yes 
    & \cellcolor{red} No
    & \cellcolor{green}Yes
    & \cellcolor{red} No
    & \cellcolor{green}Yes\\
& Validation 
	& \cellcolor{green}Yes	
    & \cellcolor{green}Yes \footnote{\label{tdd}via TDD} 
    & \cellcolor{green}Yes \footnote{\label{weeklyDemo}via weekly demos} 
    & \cellcolor{green}Yes 
    & \cellcolor{green}Yes
    & \cellcolor{green}Yes
    & \cellcolor{green}Yes\\
& Quick validation 
	& \cellcolor{red} No
    & \cellcolor{green}Yes \footnoteref{tdd} 
    & \cellcolor{green}Yes \footnoteref{weeklyDemo} 
    & \cellcolor{green}Yes
    & \cellcolor{green}Yes
    & \cellcolor{green}Yes
    & \cellcolor{green}Yes\\
& Focus on customer 
	& \cellcolor{red} No 
    & \cellcolor{green}Yes \footnote{via AMDD} 
    & \cellcolor{green}Yes 
    & \cellcolor{red} No
    & \cellcolor{green}Yes
    & \cellcolor{green}Yes
    & \cellcolor{red}No\\
& Understandability guarantee 
	& \cellcolor{green}Yes \footnote{Accomplished with lengthy documentation and planning.}
    & \cellcolor{red}No 
    &\cellcolor{red}No 
    &\cellcolor{red}No
    & \cellcolor{red}No
    & \cellcolor{green}Yes \footnote{Via continuous user engagement in the development process}
    & \cellcolor{red}No\\
& Technical debt control 
	& \cellcolor{green}Yes
    & \cellcolor{red}No 
    &\cellcolor{red}No 
    & \cellcolor{green}Yes
    & \cellcolor{red}No
    & \cellcolor{red}No
    & \cellcolor{red}No\\
\hline
\hline
 \multirow{7}{*}{\rotatebox{90}{Agility Criteria}}
& Prioritizes added value
	& \cellcolor{red}No
    & \cellcolor{green}Yes 
    & \cellcolor{green}Yes
    & \cellcolor{green}Yes
    & \cellcolor{green}Yes
    & \cellcolor{green}Yes
    & \cellcolor{green}Yes\\
& Allows partial requirements
	& \cellcolor{red}No	
    & \cellcolor{green}Yes 
    & \cellcolor{green}Yes	
    & \cellcolor{green}Yes
    & \cellcolor{green}Yes
    & \cellcolor{green}Yes
    & \cellcolor{green}Yes\\
& Focuses on small teams
	& \cellcolor{red}No	
    & \cellcolor{green}Yes 
    \footnote{Works for all sizes} 
    & \cellcolor{green}Yes	
    & \cellcolor{green}Yes
    & \cellcolor{green}Yes
    & \cellcolor{green}Yes
    & \cellcolor{green}Yes\\
& Develops minimal viable architecture 	
	& \cellcolor{red}No	
    & \cellcolor{green}Yes 
    & \cellcolor{green}Yes	
    & \cellcolor{green}Yes
    & \cellcolor{green}Yes
    & \cellcolor{green}Yes
    & \cellcolor{green}Yes\\
& Produces minimal documentation 
	& \cellcolor{red}No	
    & \cellcolor{green}Yes 	
    & \cellcolor{green}Yes	
    & \cellcolor{green}Yes
    & \cellcolor{green}Yes
    & \cellcolor{red}No
    & \cellcolor{green}Yes\\
& Relies heavily on customer feedback 	
	& \cellcolor{red}No	
    & \cellcolor{green}Yes 	
    & \cellcolor{green}Yes
    & \cellcolor{red} No
    & \cellcolor{green}Yes
    & \cellcolor{green}Yes
    & \cellcolor{red}No\\
& Susceptible to unforeseen risks 
	& \cellcolor{red}No	
    & \cellcolor{green}Yes 
    & \cellcolor{green}Yes	
    & \cellcolor{green}Yes
    & \cellcolor{red}No
    & \cellcolor{green}Yes
    & \cellcolor{green}Yes\\
\end{tabular}
\caption{Results of methodologies evaluation}
\end{center}
\end{table}

\section{Conclusion}

In this paper, we have presented a comparative analysis of a number of the most popular software development methodologies.  The results show that each of the ``agile''   methodologies has an important basic criteria that it doesn't cover, such as technical debt control or cost estimation. In smaller projects these issues can be ignored since the costs are small and the potential for technical debt is limited. For really large and complex projects, these limitations render the ``agile'' methods inapplicable. 

Therefore, we believe that these innate qualities of ``agile'' methods make them applicable only for a certain subset of the problems: small and solvable in small teams. However, the agile methods that we've looked at do not scale well. For large and sprawling projects, a paradigm that over=-emphasizes many person-to-person interactions stops working.

We find that these limitations explain the persistence of the traditional waterfall approach. Despite being bulky and prone to generating redundant documentation, it is a tried and true method that has proved to be able to deliver working products. It is understandable that many managers decide to err on the side of caution, especially considering the fact that benefits of using many agile methods are at best unclear.

We find it unsurprising that Vijayasarathy noticed over 45\% of software teams are using a hybrid approach \cite{vijayasarathy2016choice}. As we show in our table above, no one methodology properly addresses all important aspects of the software design process. By using a combination of approaches, greater results can be achieved. For example by using waterfall for project planning, scrum for short term goals, and TDD for a guarantee on software correctness, a team can mitigate the downfalls present in each method while highlighting their strengths.

\bibliographystyle{IEEEtran}
\bibliography{sample}

\begin{thebibliography}{10}
\providecommand{\url}[1]{#1}
\csname url@samestyle\endcsname
\providecommand{\newblock}{\relax}
\providecommand{\bibinfo}[2]{#2}
\providecommand{\BIBentrySTDinterwordspacing}{\spaceskip=0pt\relax}
\providecommand{\BIBentryALTinterwordstretchfactor}{4}
\providecommand{\BIBentryALTinterwordspacing}{\spaceskip=\fontdimen2\font plus
\BIBentryALTinterwordstretchfactor\fontdimen3\font minus
  \fontdimen4\font\relax}
\providecommand{\BIBforeignlanguage}[2]{{%
\expandafter\ifx\csname l@#1\endcsname\relax
\typeout{** WARNING: IEEEtran.bst: No hyphenation pattern has been}%
\typeout{** loaded for the language `#1'. Using the pattern for}%
\typeout{** the default language instead.}%
\else
\language=\csname l@#1\endcsname
\fi
#2}}
\providecommand{\BIBdecl}{\relax}
\BIBdecl

\bibitem{agilemanifesto}
\BIBentryALTinterwordspacing
{A}gile {M}anifesto. [Online]. Available: \url{http://agilemanifesto.org}
\BIBentrySTDinterwordspacing

\bibitem{turk2014limitations}
D.~Turk, R.~France, and B.~Rumpe, ``Limitations of agile software processes,''
  \emph{arXiv preprint arXiv:1409.6600}, 2014.

\bibitem{qumer2008evaluation}
A.~Qumer and B.~Henderson-Sellers, ``An evaluation of the degree of agility in
  six agile methods and its applicability for method engineering,''
  \emph{Information and software technology}, vol.~50, no.~4, pp. 280--295,
  2008.

\bibitem{martin2003agile}
R.~C. Martin, \emph{Agile software development: principles, patterns, and
  practices}.\hskip 1em plus 0.5em minus 0.4em\relax Prentice Hall PTR, 2003.

\bibitem{alliance2001agile}
A.~Alliance, ``Agile manifesto,'' \emph{Online at http://www. agilemanifesto.
  org}, vol.~6, no.~1, 2001.

\bibitem{beck2000extreme}
K.~Beck, \emph{Extreme programming explained: embrace change}.\hskip 1em plus
  0.5em minus 0.4em\relax addison-wesley professional, 2000.

\bibitem{beck1999embracing}
------, ``Embracing change with extreme programming,'' \emph{Computer},
  vol.~32, no.~10, pp. 70--77, 1999.

\bibitem{paulk2001extreme}
M.~C. Paulk, ``Extreme programming from a cmm perspective,'' \emph{IEEE
  software}, vol.~18, no.~6, pp. 19--26, 2001.

\bibitem{beck2001planning}
K.~Beck and M.~Fowler, \emph{Planning extreme programming}.\hskip 1em plus
  0.5em minus 0.4em\relax Addison-Wesley Professional, 2001.

\bibitem{brown2010enabling}
N.~Brown, R.~Nord, and I.~Ozkaya, ``Enabling agility through architecture,''
  DTIC Document, Tech. Rep., 2010.

\bibitem{kitchenham1996evaluating}
B.~A. Kitchenham, ``Evaluating software engineering methods and tool part 1:
  The evaluation context and evaluation methods,'' \emph{ACM SIGSOFT Software
  Engineering Notes}, vol.~21, no.~1, pp. 11--14, 1996.

\bibitem{sorensen1995comparison}
R.~Sorensen, ``A comparison of software development methodologies,''
  \emph{CrossTalk}, vol.~8, no.~1, pp. 10--13, 1995.

\bibitem{thayer1981major}
R.~H. Thayer, A.~B. Pyster, and R.~C. Wood, ``Major issues in software
  engineering project management,'' \emph{IEEE Transactions on Software
  Engineering}, no.~4, pp. 333--342, 1981.

\bibitem{awad2005comparison}
M.~Awad, ``A comparison between agile and traditional software development
  methodologies,'' \emph{University of Western Australia}, 2005.

\bibitem{schwaber2002agile}
K.~Schwaber and M.~Beedle, \emph{Agile software development with Scrum}.\hskip
  1em plus 0.5em minus 0.4em\relax Prentice Hall Upper Saddle River, 2002,
  vol.~1.

\bibitem{vijayasarathy2016choice}
L.~R. Vijayasarathy and C.~W. Butler, ``Choice of software development
  methodologies: Do organizational, project, and team characteristics matter?''
  \emph{IEEE Software}, vol.~33, no.~5, pp. 86--94, 2016.

\bibitem{royce1970managing}
W.~W. Royce \emph{et~al.}, ``Managing the development of large software
  systems,'' in \emph{proceedings of IEEE WESCON}, vol.~26, no.~8.\hskip 1em
  plus 0.5em minus 0.4em\relax Los Angeles, 1970, pp. 1--9.

\bibitem{benington1983production}
H.~D. Benington, ``Production of large computer programs,'' \emph{Annals of the
  History of Computing}, vol.~5, no.~4, pp. 350--361, 1983.

\bibitem{ramamoorthy1985metrics}
C.~Ramamoorthy, W.~Tsai, T.~Yamaura, and A.~Bhide, \emph{METRICS GUIDED
  METHODOLOGY.}\hskip 1em plus 0.5em minus 0.4em\relax IEEE, 1985, pp.
  111--120.

\bibitem{munassar2010comparison}
N.~M.~A. Munassar and A.~Govardhan, ``A comparison between five models of
  software engineering,'' \emph{IJCSI}, vol.~5, pp. 95--101, 2010.

\bibitem{ambler2002agile}
S.~Ambler, \emph{Agile modeling: effective practices for extreme programming
  and the unified process}.\hskip 1em plus 0.5em minus 0.4em\relax John Wiley
  \& Sons, 2002.

\bibitem{alliance2016learn}
S.~Alliance, ``Learn about scrum,'' \emph{Scrum Alliance}, 2016.

\bibitem{janzen2005test}
D.~Janzen and H.~Saiedian, ``Test-driven development concepts, taxonomy, and
  future direction,'' \emph{Computer}, vol.~38, no.~9, pp. 43--50, 2005.

\bibitem{damm2005introducing}
L.-O. Damm, L.~Lundberg, and D.~Olsson, ``Introducing test automation and
  test-driven development: An experience report,'' \emph{Electronic Notes in
  Theoretical Computer Science}, vol. 116, pp. 3--15, 2005.

\bibitem{maximilien2003assessing}
E.~M. Maximilien and L.~Williams, ``Assessing test-driven development at ibm,''
  in \emph{Software Engineering, 2003. Proceedings. 25th International
  Conference on}.\hskip 1em plus 0.5em minus 0.4em\relax IEEE, 2003, pp.
  564--569.

\bibitem{bissi2016effects}
W.~Bissi, A.~G. S.~S. Neto, and M.~C. F.~P. Emer, ``The effects of test driven
  development on internal quality, external quality and productivity: A
  systematic review,'' \emph{Information and Software Technology}, vol.~74, pp.
  45--54, 2016.

\bibitem{martin1991rapid}
J.~Martin, \emph{Rapid application development}.\hskip 1em plus 0.5em minus
  0.4em\relax Macmillan Publishing Co., Inc., 1991.

\bibitem{jacobson1999unified}
I.~Jacobson, G.~Booch, J.~Rumbaugh, J.~Rumbaugh, and G.~Booch, \emph{The
  unified software development process}.\hskip 1em plus 0.5em minus 0.4em\relax
  Addison-wesley Reading, 1999, vol.~1.

\bibitem{gerber2007practical}
A.~Gerber, A.~Van Der~Merwe, and R.~Alberts, ``Practical implications of rapid
  development methodologies.''\hskip 1em plus 0.5em minus 0.4em\relax Computer
  Science and Information Technology Education Conference, 2007.

\bibitem{beynon1999rapid}
P.~Beynon-Davies, C.~Carne, H.~Mackay, and D.~Tudhope, ``Rapid application
  development (rad): an empirical review,'' \emph{European Journal of
  Information Systems}, vol.~8, no.~3, pp. 211--223, 1999.

\bibitem{mackay2000reconfiguring}
H.~Mackay, C.~Carne, P.~Beynon-Davies, and D.~Tudhope, ``Reconfiguring the
  user: using rapid application development,'' \emph{Social studies of
  science}, vol.~30, no.~5, pp. 737--757, 2000.

\bibitem{carmel1993pd}
E.~Carmel, R.~D. Whitaker, and J.~F. George, ``Pd and joint application design:
  a transatlantic comparison,'' \emph{Communications of the ACM}, vol.~36,
  no.~6, pp. 40--48, 1993.

\bibitem{hirschheim1991symbolism}
R.~Hirschheim and M.~Newman, ``Symbolism and information systems development:
  myth, metaphor and magic,'' \emph{Information Systems Research}, vol.~2,
  no.~1, pp. 29--62, 1991.

\bibitem{ives1984user}
B.~Ives and M.~H. Olson, ``User involvement and mis success: A review of
  research,'' \emph{Management science}, vol.~30, no.~5, pp. 586--603, 1984.

\bibitem{davidson1993exploratory}
E.~J. Davidson, \emph{An exploratory study of joint application design (JAD) in
  information systems delivery}.\hskip 1em plus 0.5em minus 0.4em\relax Center
  for Information Systems Research, Sloan School of Management, Massachusetts
  Institute of Technology, 1993.

\bibitem{becker1993integrating}
S.~A. Becker, E.~Carmel, and A.~R. Hevner, ``Integrating joint application
  development (jad) into cleanroom development with icase,'' in \emph{System
  Sciences, 1993, Proceeding of the Twenty-Sixth Hawaii International
  Conference on}, vol.~3.\hskip 1em plus 0.5em minus 0.4em\relax IEEE, 1993,
  pp. 13--21.

\bibitem{hunt2006feature}
J.~Hunt, ``Feature-driven development,'' \emph{Agile Software Construction},
  pp. 161--182, 2006.

\bibitem{benoit1999feature}
M.~Benoit, R.~Anthony, and L.~B. Wee, ``Feature-driven development,'' 1999.

\bibitem{palmer2001practical}
S.~R. Palmer and M.~Felsing, \emph{A practical guide to feature-driven
  development}.\hskip 1em plus 0.5em minus 0.4em\relax Pearson Education, 2001.

\bibitem{khramtchenko2004comparing}
S.~Khramtchenko, ``Comparing extreme programming and feature driven development
  in academic and regulated environments,'' \emph{Feature Driven Development},
  2004.

\end{thebibliography}

\end{document}